\documentclass[12pt]{iopart}
\usepackage{iopams}
\usepackage[xdvi]{graphicx}
\usepackage{slashbox}

\newcommand{\bra}{\langle}
\newcommand{\ket}{\rangle}

\newcommand{\der}[2]{\frac{d #1}{d  #2}}

\newcommand{\bv}[1]{{\boldsymbol #1}}

\newcommand{\muc}{\mu_{\rm c}}

\begin{document}

%\hfill {\bf ver. 9  11/05/24}

% Journal identifier can be put here if required, e.g.
% \jl{14}

\title
[Statistical mechanics of glass transition in 
lattice molecule models]
{Statistical mechanics of glass transition in 
lattice molecule models}

\author{Shin-ichi Sasa
\footnote[3]
{sasa@jiro.c.u-tokyo.ac.jp}}

\address{Department of Pure and Applied Sciences, 
University of Tokyo, Komaba, Tokyo 153-8902, Japan}

\begin{abstract}
Lattice molecule models are proposed in order to 
study statistical mechanics of glass transition
in finite dimensions. Molecules in the models 
are represented by hard Wang tiles and their 
density is controlled by a chemical potential. 
An infinite series of irregular ground states 
are constructed theoretically. By defining 
a glass order parameter as a collection of 
the overlap with each ground state, 
a thermodynamic transition to a glass phase 
is found in a stratified Wang tiles model 
on a cubic lattice.
\end{abstract} 


% 05.60.+q Lattice theory and statistics
% 64.60.De Statistical mechanics of model systems
% 75.10.Hk Classical spin models

% \submitted

%%%%%%%%%%%%%% main text starts %%%%%%%%%%%%%%%%%%%
\section{Introduction}

%% status of glass transition

The equilibrium statistical mechanics successfully 
describes various types of phase transitions including
ferromagnetic-paramagnetic transitions, gas-liquid 
transitions, and  liquid-solid transitions. For structural
glass transitions, whose precise definition is not obvious, 
the understanding has been accumulated from several 
viewpoints \cite{Cavagna}. 
In particular, in addition to an insightful 
phenomenological argument 
\cite{GibbsDiMarzio,AdamGibbs,KTW,BB:JCP}, 
which is often referred
to as {\it a random first-order transition scenario} (RFOT),
the analysis within the equilibrium statistical mechanics 
has provided  quantitative results for structural 
glass transitions \cite{Monasson,MezardParisi,
BiroliMezard,Rivoire,Krzakala,ParisiZamponi}. 
At present, it has become widely conjectured 
that a thermodynamic glass transition, if it 
exists, is described as a one-step replica 
symmetry breaking (1RSB) in the  spin glass terminology. 
Despite such successes, a theory of glass 
transition in finite-dimensional, short-range interaction
systems is still one of 
challenging problems in physics, because the theory 
on the basis of equilibrium statistical mechanics
has been  established only for models with infinite-range 
interaction or on a random graph. 

%% critical phenomena

Here, let us recall a history of studies on critical
phenomena. The van der Waals theory is the first breakthrough 
of theory for critical phenomena, which might 
correspond to RFOT for glass transitions. It should be noted
that the statistical mechanics of a model with an infinite-range 
interaction exactly predicts critical phenomena in 
accordance with the van der Waals theory \cite{Gallavotti}. 
This reminds us a relation between 1RSB in  mean-field type 
models and RFOT in phenomenology. 
Then, the existence of a critical point within 
the equilibrium statistical mechanics of short-range interaction
Hamiltonians was shown by Peierls \cite{Peierls}, 
Kramers-Wannier \cite{KrameresWannier}, 
and Onsager \cite{Onsager}.
In particular, Onsager solved the two-dimension Ising model exactly
and proved that critical phenomena in finite dimensions 
are qualitatively different from  the van der Waals 
theory (or the Curie-Weiss theory in ferromagnetic-paramagnetic 
transitions). 
Since then, the significance of critical phenomena in 
finite dimensions has been recognized and much effort has
been done in order to connect between Onsager's result 
and the van der Waals theory.
We now understand a great picture of critical phenomena.

%% status of studies 

However, with regard to glass transitions, there is  no exactly 
solved example in finite dimensions; rather, there is no finite-dimensional,
short-range interaction model for which the existence of a glass 
transition is understood 
theoretically.  Thus far, toward  establishment of statistical
mechanics of glass transition in finite dimensions, 
several lattice models  have been proposed. 
One example is a class of models proposed by Biroli 
and M\'ezard \cite{BiroliMezard}, 
in which at most $k$ neighboring particles are allowed to 
contact with each particle. Although this simple model 
exhibits a glass transition when it is defined on a  random graph,
crystallization occurs in  finite-dimensional lattices 
except for  subtle cases \cite{HukushimaSasa}. 
In the other model proposed by Ref. \cite{LatticeGlass},
crystallization might be prohibited, and the numerical experiment 
has been performed in order to explore the nature of thermodynamic
glass transitions \cite{Parisi}. However, it seems difficult to
develop a theoretical argument for this model in finite dimensions. 
Furthermore, finite-dimensional quenched-disordered spin  models  
that exhibit 1RSB  under the mean-field approximation 
have been studied 
numerically \cite{disorder-spin}. However, the numerical computation
is  much harder than standard spin glass models, and a precise theory 
for the model might be quite challenging. (See Ref. \cite{Moore} as such
a theoretical attempt.)

% our strategy - basic idea : order parameter 

In contrast to  previous studies, we first  propose 
finite-dimensional hard-constraint models  for which ground states
can be constructed theoretically. Here,  the ground states are 
obtained by taking the limit of the chemical potential to be 
infinitely large, because  the chemical potential is the only 
thermodynamic intensive variable in such models. Indeed, we can 
show that our model possesses uncountably infinite number of
ground states in the infinite size lattice, and  typical 
ground states are irregular
in the sense that they do not exhibit any long-range positional 
order characterized by the existence of Bragg peaks. Note that 
quasi-periodic ground states are classified as regular ground states.
Let us denote a set of all ground-state configurations 
by ${\cal D}$, which we can specify theoretically for our models. 
Now, we introduce  an overlap with a  ground state 
configuration $\alpha \in {\cal D}$, which is denoted by
$q_{\alpha}$. Since $q_{\alpha}$ is defined for each ground 
state $\alpha$, we have  an infinite-dimensional vector 
$\bv{q}=(q_{\alpha})_{\alpha \in {\cal D}}$.
We refer this vector to as {\it the  order parameter},
because this corresponds to the magnetization in the Ising model.

% our strategy - basic idea : GS-boundary conditions

We explain the correspondence by reviewing 
the phase transition in the two-dimensional 
Ising model. When the temperature is higher than the critical 
temperature, in the thermodynamic limit, there exists a unique 
expectation value  of observables with being independent of boundary 
conditions. However, the independence of boundary conditions is broken 
below the critical temperature. In general, a state of the system 
without the uniqueness is referred to as  {\it the ordered phase}.
The dependence on boundary conditions is most easily 
observed when the expectation of the  magnetization is considered
under the spin-up boundary condition or the spin-down boundary 
condition. We here notice that the magnetization is equivalent
to the overlap with one  ground state. We thus consider the
overlap with each ground state as the generalization of the
magnetization. We  also generalize
spin-up and  spin-down boundary conditions to special 
boundary conditions that uniquely determine a ground-state 
configuration for a series of system sizes going to the 
infinity. We call such a boundary 
condition a {\it GS-boundary condition}. 

% The purpose of our paper and what we did

Since we have the glass order parameter $\bv{q}$ in our models, 
we can investigate whether or not the expectation value of 
$\bv{q}$ takes a different value under every GS-boundary condition.
If the dependence is shown,  the existence of an  ordered 
phase characterized by the parameter $\bv{q}$ is claimed.  
From a fact that a typical  ground state does not  exhibit
a long-range positional order, we identify the ordered phase 
as {\it the glass phase}.  
As far as we know, such an approach to glass problems  
has never been attempted. 

% road map

This paper is organized as follows.
In section \ref{model-wang}, we start with a definition of
lattice molecule models we study.
The molecules in the models are represented by hard Wang tiles 
\cite{Tiling} and the molecule density obeys 
a grand canonical ensemble. 
In section \ref{model-4}, 
we consider a simple  model in this class. We show 
that the model possesses an infinite series of irregular 
ground-states, while no thermodynamic 
transition occur. In section 
\ref{3dmodel}, as an extension of the model, 
we propose  a three-dimensional model in which a
thermodynamic glass transition is observed. 
We present theoretical arguments 
and numerical evidences for the thermodynamic glass transition.
Section \ref{remarks} is devoted to concluding remarks.

%%%%%%%%%%%%%%%%%%%%%%%%%%%%%%%%%%%%%%%%%%%%%%%%%%%%%%%
\section{Lattice molecule model}\label{model-wang}    %
%%%%%%%%%%%%%%%%%%%%%%%%%%%%%%%%%%%%%%%%%%%%%%%%%%%%%%%

\begin{figure}[htbp]
\begin{center}
\includegraphics[width=10cm]{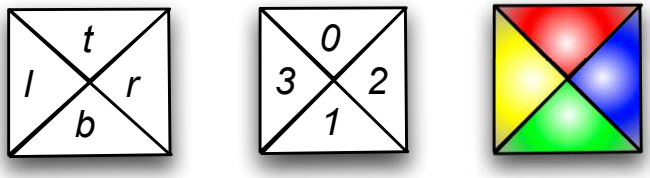}
\end{center}
\caption{Unit square tile with colored edges.}
\label{tile}
\end{figure} 

% Wang tiles 

Let $\Lambda_L=\{ (i,j) \in {\bf N}\times {\bf N}| 1 \le i,j \le L \}$
be a square lattice. We formulate a statistical mechanical model in 
the lattice $\Lambda_L$. Each site can be occupied by at most one 
molecule. A  molecule is characterized by its shape, represented by 
$m$-colors given on edges of a unit square. 
Since  tiles with colored edges are called  {\it Wang tiles},
the molecules in our model may be interpreted as Wang tiles.

% Coloring

When left, right, bottom and top  edges of a  Wang tile are 
colored by $\ell$, $r$, $b$ and $t$, respectively, we denote 
the  quartet of colors by $(\ell,r,b,t)$.
Below, the colors will be identified with integers if the 
correspondence is explicitly given. (See figure \ref{tile}.)  
Among  $m^4$ different Wang tiles, we select $p$ different 
tiles. These are called {\it prototiles} and the set of prototiles
is denoted by $C$. Each  prototile is  represented  by an 
integer $n$, $1 \le n \le p$.  At each site $x \in \Lambda_L$,
we define $\sigma(x)=n$ if there is a molecule
congruent to a prototile $n \in C$  and as $\sigma(x)=0$ 
if the site is empty. We call an empty site a {\it hole}. 
The set of tile configurations $(C \cup \{0\})^{\Lambda_L}$ 
is denoted by $\Sigma_L$. 
We study  statistical mechanics of the  molecules under a boundary 
condition imposed at sites in $\partial \Lambda_L^-$ and
$\partial \Lambda_L^+$, where 
$\partial \Lambda_L^-
\equiv \{(i,j) \in \Lambda_L |i=1 \ {\rm or}\  j=1\} $
and 
$\partial \Lambda_L^+ \equiv \{(i,j) \in \Lambda_L |i=L \ {\rm or } \ j=L\} 
$. The bulk region $ \Lambda_L \backslash
(\partial \Lambda_L^+ \cup \partial \Lambda_L^-) $ 
is denoted by $\bar \Lambda_L$, and 
the number of sites in the bulk region is $N=(L-2)^2$. 

% hard interaction

Specifically, we consider the case  that the interaction 
between molecules is described by a hard constraint that molecules 
are allowed to contact each other only when contiguous edges 
of tiles have the same color. We then assume a grand canonical 
ensemble
\begin{equation}
P_\mu(\sigma)=\frac{1}{\Xi(\mu)}D(\sigma)e^{\mu N \rho(\sigma)},
\label{gc}
\end{equation}
where $D(\sigma)=1$ for configurations that satisfy the constraint
that contiguous edges have the same color, otherwise $D(\sigma)=0$;
$\mu$ is the chemical potential of molecules taking a value 
in $[-\infty, \infty]$, 
and $\rho(\sigma)$ is the density of molecules defined by 
\begin{equation}
\rho(\sigma)=\frac{1}{N} \sum_{x \in \bar \Lambda_L}[1-\delta(\sigma(x),0)].
\label{density}
\end{equation}
Here, the temperature is set to be unity and its value is irrelevant 
for the problem. The normalization constant $\Xi(\mu)$ in (\ref{gc})
is the partition function, which is explicitly given by
\begin{equation}
\Xi(\mu)=\sum_{\sigma \in \Sigma_L}D(\sigma)e^{\mu N \rho(\sigma)}.
\label{Xi}
\end{equation}
Configurations realized in the limit $\mu \to \infty$ are 
ground-states in statistical mechanics. A  tile configuration 
without holes, which is referred to as a {\it complete tiling} 
in this paper, provides a ground state.

% statistical mechanics

% general formalism 

Thermodynamic properties associated with the density 
are determined by the pressure function $p(\mu)$ defined by 
\begin{equation}
p(\mu)\equiv \lim_{N \to \infty} \frac{1}{N}\log \Xi(\mu).
\label{pressure}
\end{equation}
The expectation value of the density $\bra \rho \ket$
is given by 
\begin{equation}
\bra \rho \ket=\der{p(\mu)}{\mu}.
\end{equation}
Furthermore, the entropy density $s$ is 
related to the pressure in term of the Legendre transformation
\begin{equation}
s(\rho)=\inf_{\mu} [p(\mu)-\rho \mu].
\label{entoropy}
\end{equation}

% example (Ising)

Statistical behavior  of the model depends on the choice of 
a set of $p$-types of molecules $C$. As the simplest example, 
let us consider the set $C$ with $p=2$, in which all the edges 
of one type  are red and all the edges of the other type 
are green. In this model, ground states are  
understood as the complete tilings occupied by one color 
when the open boundary condition 
$\sigma(x)=0$ for $x \in \partial \Lambda_L^- \cup 
\partial \Lambda_L^+$ is assumed. 
When $\mu$ is sufficiently large, the number 
density of red tiles depends on boundary condition 
even in the limit $L \to \infty$.  On the other hand, 
when $\mu$ is sufficiently small (negatively large), 
tile configurations 
are disordered and all statistical quantities are 
independent of  boundary conditions in the limit $L \to \infty$.  
The ${\bf Z}_2$ symmetry breaking occurs at some $\mu$ beyond 
which there exists an ordered  phase. The universality 
class near the transition is identical to that of the 
two-dimensional Ising model. 

% aperiodic complete tiling

A unique feature of Wang-tiles is that the operation of any 
Turing machine is simulated by a complete tiling for an appropriate 
set $C$. (See Ref. \cite{Tiling} for the research history.
See also  Ref. \cite{Robinson} as an instructive paper for 
this issue.) According to  computation theory, 
this means that there is no algorithm  that determines 
whether or not a complete tiling is possible for a given 
set $C$ \cite{Davis}. In solving the decidability problem,
it was a crucial step to find {\it an aperiodic set of 
prototiles} $C$ for which an aperiodic complete tiling in 
$\Lambda_\infty$ exists, while periodic complete tilings 
cannot be realized. After the first discovery, $p$, the 
number of elements of an aperiodic set has been reduced. 
At present, the minimum number of $p$ is 13 \cite{Culik}. 
Statistical mechanics of a system consisting of 
an aperiodic set of 16-prototiles was studied 
in Ref. \cite{Leuzzi}, where holes are not considered, 
but a positive energy is assumed  for mismatches of 
contiguous colors. This reference claims that a 
thermodynamic transition occurs at some finite temperature. 
See also Ref. \cite{Radin} for the recent study on the
model.

%  statisitical mechanics
% in the sense of dynamical systems

Here, from a viewpoint of statistical mechanics, 
it is important to recognize that there exits a set $C$
with which non-trivial ground states can be obtained
as complete tilings corresponding to computational 
processes. We do not need to stick to aperiodic
sets of Wang tiles. More important thing in the 
context of glass problems is that ground states
should not possess any long-range positional order. However,
in the construction of complete tilings with using 
the aperiodic sets with $p=13$ and $p=14$ \cite{Kari},  the 
quasi-periodic maps are employed to yield the tiling \cite{cha},
and therefore their complete tilings possess the 
quasi-periodic order. The construction method in 
the other cases are entirely different from the cases 
that $p=13$ and $14$, but at least for known examples 
in Ref. \cite{Tiling}, the  complete tilings seem to
exhibit a long-range positional (quasi-periodic) order. 

Now, let us recall a dynamical-system theory,
which tells us that aperiodic motion described as a solution 
of a deterministic equation is  further classified into 
quasi-periodic motion and  irregular motion \cite{DS,ER}. 
Periodic and quasi-periodic motion are called 
{\it regular motion} and characterized by the existence 
of the singular peak in its spectrum. Then, there are an 
infinitely number of periodic orbits in typical chaotic 
systems and the exclusion of periodic orbits can be 
realized by systems that exhibit quasi-periodic motion. 
Similarly, it is reasonable 
to classify aperiodic configurations generated by a deterministic 
rule into quasi-periodic and irregular configurations. Quasi-periodic 
configurations are characterized by the existence of a singular 
peak in the  Fourier transform (Bragg peak) of some representation
of configuration, as is known in quasi-crystals \cite{QC}.
See also Ref. \cite{Enter1} for a mathematical argument
of the definition of {\it weak crystals} which cover certain
generalization of quasi-crystals. In our viewpoint, non-periodic 
long-range positional order in Thue-Morse sequences \cite{Enter2}
is classified into  the same group as quasi-periodic order. Here,
it should be noted that irregular configurations 
without any Bragg peaks can also be  generated by 
a deterministic rule. In order to seek for thermodynamic glass 
transitions, we study statistical mechanics associated with such 
irregular (no periodic and no quasi-periodic) complete tilings
in $\Lambda_\infty$. 

From this reason, 
we are not concerned with aperiodic sets of prototiles,
but with the case that complete tilings are irregular 
almost surely when a complete tiling is picked up with 
the equal weight, while there exists a countably 
infinite number of  periodic complete tilings. 
We shall provide an example in the next section. 

%%%%%%%%%%%%%%%%%%%%%%%%%%%%%%%%%%%%%%%%
\section{4-prototiles model}\label{model-4} %
%%%%%%%%%%%%%%%%%%%%%%%%%%%%%%%%%%%%%%%%

\begin{figure}[htbp]
\begin{center}
\includegraphics[width=8cm]{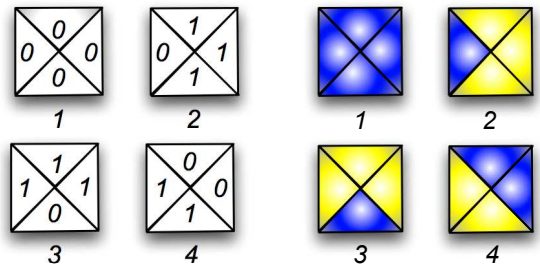}
\end{center}
\caption{Set of 4-prototiles. The left figure shows 
a number representation. The tiles satisfy 
$t=r=\ell+b$ (mod 2). The right 
figure shows a color representation.
}
\label{4tiles}
\end{figure} 

% model

The simplest set of Wang prototiles that generates 
irregular complete tilings is given in figure 
\ref{4tiles}. This set  is characterized  by a rule 
that the quartet $(\ell,r,b,t)$  satisfies 
$t=r=\ell+b$ (mod 2), 
where $r$, $\ell$, $b$ and $t$ are either 0 or 1.
For this model, we shall show that (i) there are 
an uncountably infinite number of irregular 
complete tilings, but (ii) no thermodynamic 
transition occurs. 

\subsection{Complete tilings}\label{complete}

% CA-construction of densest configurations

\begin{table}[htbp]
\begin{center}
\begin{tabular}{|c c||c|c|}
\hline
\hline
2  3 &1& 3 & 4   \\
\hline
1  4 &0& 1 & 2 \\
\hline
\hline
&\slashbox{$r$}{$t$} & 0  &1 \\ 
\slashbox{$\leftarrow$}{$\downarrow$}  & &1 4  &  2 3  \\
\hline
\end{tabular} 
\caption{
Cellular automaton rule for generating complete tilings
in $\Lambda_{\infty}$. 
When $r$  of the left tile and $t$ 
of the bottom tile are given, the
tile is uniquely determined. The rule is equivalent to 
Rule 102 in Ref. \cite{CA}.}
\label{carule}
\end{center}
\end{table}

% coding 

Complete tilings in this model are obtained as follows.
For an element $([h(i)]_{i=1}^L,[v(j)]_{j=1}^L )$
in the set ${\cal D}_L\equiv \{0,1 \}^L \times \{0,1\}^L$, 
we  set $\ell(1,j)=v(j)$ and $b(i,1)=h(i)$. Then, by
the rule in table \ref{carule}, tiles at $(i,1)$, $2 \le i \le L$, 
are determined from smaller $i$ in order. 
Similarly, the tiles at $(i,j)$, $1 \le i \le L$, are 
determined for $j=2,3,\cdots, L$. In this manner, all 
the complete tilings in $\Lambda_L$ are uniquely coded by 
${\cal D}_L$.  That is, there are $2^{2L}$ complete
tilings for the system of size $L$. 
The complete tilings in $\Lambda_\infty$ are obtained
in the limit $L \to \infty$.
Since  ${\cal D}_\infty$ 
has one-to-one correspondence with real numbers in the interval 
$[0,1]$, the cardinality of the complete tilings in $\Lambda_\infty$
are uncountably infinite. 
Since periodic tilings are countable, almost all complete 
tilings are aperiodic.  

% irregularity 

\begin{figure}[htbp]
\begin{center}
\includegraphics[width=6cm]{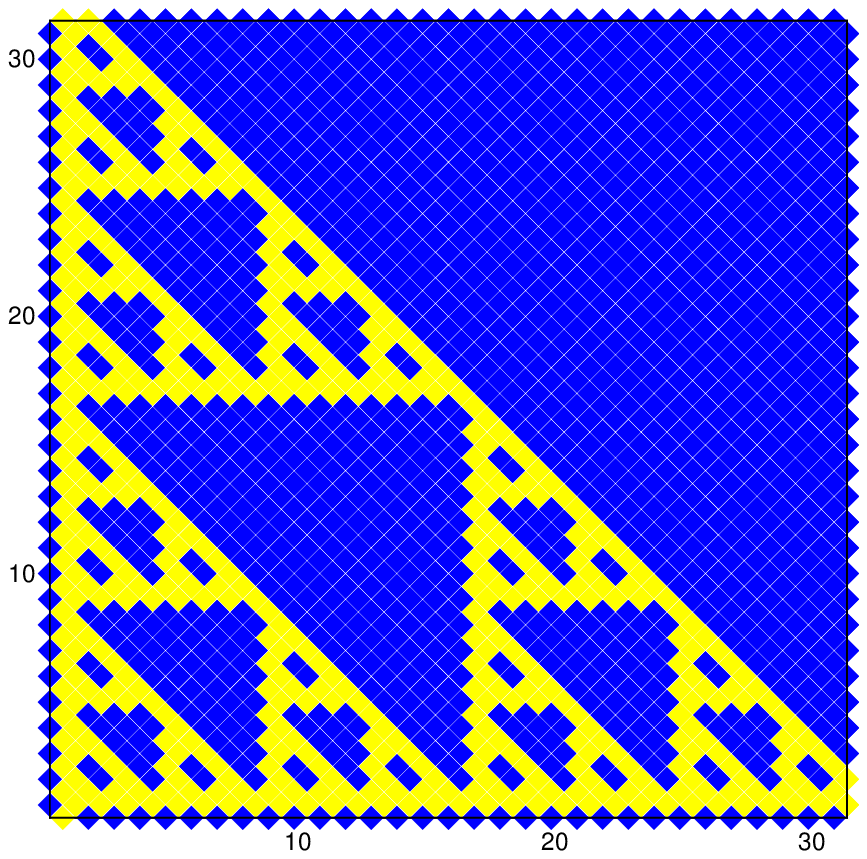}
\includegraphics[width=6cm]{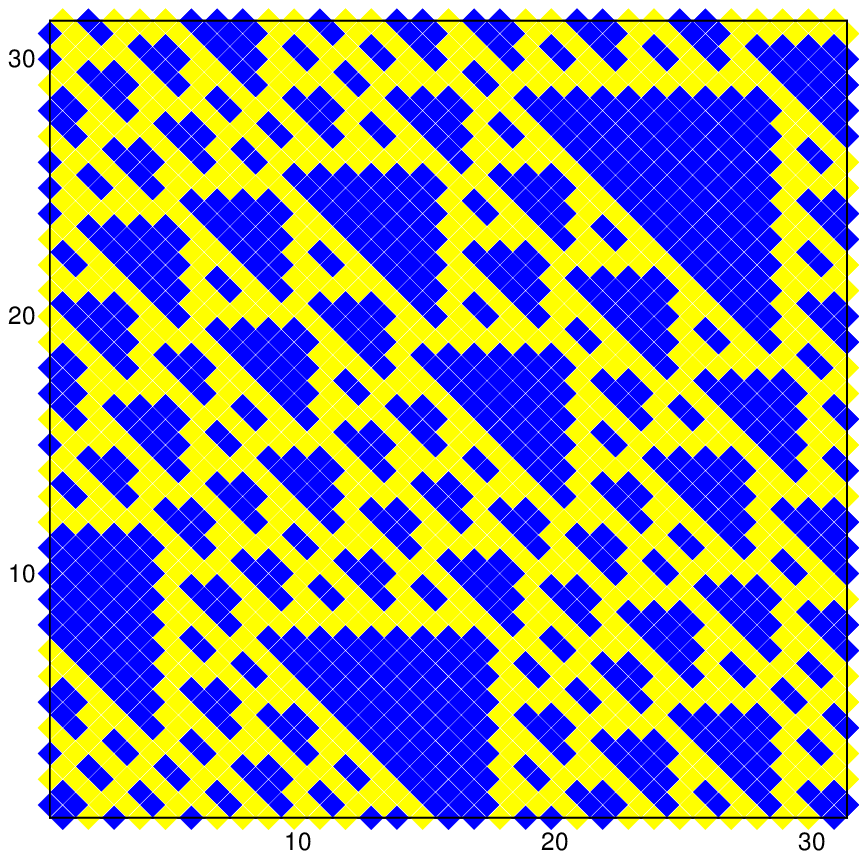}
\end{center}
\caption{Examples of complete tilings. $L=32$. 
Left: $h(i)=\delta(i,1)$ and $v(j)=0$. 
Right: $h(i)$ and $v(j)$ are chosen randomly.
}
\label{tilepattern}
\end{figure}

A remarkable feature of the complete tilings is the additivity.
Suppose that $\sigma_1$ and $\sigma_2$ are different complete 
tilings, respectively. Here, we define the addition of two 
configurations $\sigma_1$ and $\sigma_2$ by determining
$\ell(x)$ and $b(x)$ as
\begin{eqnarray}
\ell(x) &=& \ell_1(x)+\ell_2(x) \quad ({\rm mod} \quad 2), \\
b(x)&=&b_1(x)+b_2(x) \quad ({\rm mod} \quad 2)
\end{eqnarray}
at each site $x \in \Lambda_L$. 
We denote this addition by $\sigma_1 \oplus \sigma_2$. 
Then, from the coloring  rule in figure \ref{4tiles}, 
we find that the configuration 
$\sigma_1 \oplus \sigma_2$ is also another complete tiling.  
Note that the configuration  $\sigma(x)=1$ for all 
$x \in \Lambda_L$  is one of complete tilings. This tiling, 
which is denoted by $\sigma_{0,0}$, is the unit element in 
the additive group consisting of all complete tilings.
As the simplest complete tiling other than $\sigma_{0,0}$, 
we generate a configuration from a condition 
that $h(i_0)=1$ and $h(i)=0$ for $i \not = i_0$ and
$v(j)=0$ for any $j$. As displayed in the  left-side of figure 
\ref{tilepattern},  a fractal pattern, which is 
known as a Sierpinski gasket, is obtained.
We denoted it by $\sigma_{i_0,0}$. Similarly, we can generate
a  fractal configuration $\sigma_{0,j_0}$  from 
$v(j_0)=1$ and $v(j)=0$ for $j \not = j_0$ and
$h(i)=0$ for any $i$. All complete tilings in $\Lambda_\infty$
are given 
by a superposition of these basic configurations as
\begin{equation}
\sum_{i,n=1}^\infty\hspace{-1.5em}\bigcirc\hspace{0.5em}
[h(i) \sigma_{i,0} \oplus v(j)\sigma_{0,j}] ,
\label{super}
\end{equation}
where $\sum\hspace{-1.0em}\circ\hspace{0.5em}$ 
represents the summation of the configurations
in the sense of $\oplus$. When we uniformly choose one 
complete tiling, it is given by a random superposition 
of Sierpinski gaskets, as displayed in the right side of figure 
\ref{tilepattern}.
The randomness of $h(i)$ and $v(j)$ in (\ref{super}) yields
irregular configurations.

%%% complete-tling (fractal) and random superposition

% landscape

As a preliminary for later arguments, we define 
the overlap ratio between two complete tilings 
$\alpha, \alpha' \in {\cal D}_\infty$:
\begin{equation}
q_{\alpha,\alpha'}
\equiv \frac{1}{L^2}
\sum_{x \in \Lambda_L}\delta(\sigma_\alpha(x) ,\sigma_{\alpha}(x) ),
\end{equation}
and we consider the probability density $P_{\rm GS}(q)$ 
that $q_{\alpha,\alpha'}$ takes a value $q$ when 
$\alpha$ and $\alpha'$ are sampled uniformly.
Since the 4-prototiles in our model are  distributed with 
an equal ratio for a typical sample uniformly chosen
from ${\cal D}_\infty$, we have $P_{\rm GS}(q) = \delta(q-1/4)$
in the limit $L \to \infty$.

%% metion tmy previous work

At the end of section \ref{complete}, 
let us remark  previous studies related to the 4-prototiles 
model. The complete tilings of the model  are 
essentially the same as the ground-state configurations 
of a three-body interaction spin model defined on upward 
triangles \cite{NM,GN,JG}. While an extremely slow dynamics
is observed, no thermodynamic phase transition occur at a 
finite temperature. Recently, it has been shown that the system 
under a magnetic field, which is equivalent to a lattice 
gas model with a chemical potential, exhibits a first-order 
transition in the $(T,\mu)$ space \cite{Sasa}. 
Although  non-trivial nature of the coexistence 
phase has been conjectured, further studies remain to be done.

\subsection{Statistical Mechanics}\label{SM-model4}

% hole variable 

We consider statistical mechanics of the 4-prototiles model.
It is convenient to introduce a hole variable 
\begin{equation}
\eta(x) \equiv \delta(\sigma(x),0).
\label{hole-def}
\end{equation}
That is,  $\eta(x)=1$ only if the  site $x$ is empty.
For a given hole configuration  $\eta \in \{0,1\}^N $, 
we define a region 
$N_{\eta} \equiv \{ x \in \Lambda_L | \eta(x) =0 \}$ 
in which tiles occupy the sites. Let $\tilde \sigma: 
N_\eta \to C$ be a restriction of $\sigma$ on the region 
$N_\eta$. Any 
configuration $\sigma$ can be then expressed by variables 
$\eta$ and $\tilde \sigma$. The partition function $\Xi(\mu)$
is expressed as 
\begin{equation}
\Xi(\mu)=\e^{\mu N} \sum_{\eta \in \{0, 1\}^N } 
\e^{-\mu\sum_{x \in \bar \Lambda_L} \eta(x)} 
\Omega(\eta),
\label{eta-Xi}
\end{equation}
where $\Omega(\eta)$ is the number of 
possible tile configurations $\tilde \sigma$ when a hole 
configuration $\eta$ is given. We here fix one complete tiling 
$\alpha$ in $\Lambda_\infty$.
We assume a boundary condition that 
$\sigma(x)=\sigma_\alpha(x)$
for $x  \in \Lambda_L^-$ 
and $\sigma(x)=0$ for $ x \in \Lambda_L^+$.
We can calculate $\Omega(\eta)$ most easily 
under the boundary condition. 
Note that the pressure function $p(\mu)$ in the thermodynamic
limit does not depend on the choice of boundary conditions.

%% density of states for a given  hole configurations

% start

The number $\Omega(\eta)$ is estimated as follows.
It is obvious that $\Omega(\eta)=1$ when there are no holes. 
Suppose that there is a hole at the  site $(i_1,n_1)$.
Then, two tiles are possible at the site $(i_1+1,n_1)$
if there is no hole at the site $(i_1+1,n_1)$. We choose one tile of 
the two. For each choice, another tile configuration  is uniquely 
determined  by repeating the rule in Table \ref{carule} with
starting from the site $(i_1+1,n_1)$. Note that the configurations 
are consistent with the boundary conditions at $i=L$ and $n=L$. 
Similarly, there are two tile configurations that 
originate from the tile replaced at the site $(i_1,n_1+1)$
if there is no hole at the site $(i_1,n_1+1)$. 
Therefore, the number of possible configurations is
\begin{equation}
2^{1-\eta(i_1+1,n_1)}\cdot  2^{1-\eta(i_1,n_1+1)},
\label{basic}
\end{equation}
where we set $\eta(L,n)=\eta(i,L)=1$ in accordance with
the boundary conditions.

% second - independence

Next, we assume that there is another hole at the  site $(i_2,n_2)$.
Then, in a manner similar to the first case, the number of 
possible tile configurations is assigned to this hole.
The important thing here is that this estimation is obtained 
with being independent of configurations generated 
from the first hole at the site $(i_1,n_1)$. Since 
there is no over-counting of configurations, the number of
possible configurations is
\begin{equation}
2^{1-\eta(i_1+1,n_1)} \cdot 2^{1-\eta(i_1,n_1+1)}
\cdot 2^{1-\eta(i_2+1,n_2)} \cdot 2^{1-\eta(i_2,n_2+1)}.
\end{equation}
Repeating these considerations 
for all holes, we arrive at the expression 
\begin{equation}
\Omega(\eta)
= \prod_{i,n} 
2^{\eta(i,n)(1-\eta(i+1,n))} 2^{\eta(i,n)(1-\eta(i,n+1))}.
\label{Omega4}
\end{equation}
By introducing  a spin variable $s(x)\equiv 2\eta(x)-1$,
we rewrite (\ref{Omega4}) as 
\begin{equation}
\Omega(\eta)
= e^{\frac{J}{2}N- H(s)},
\label{Omega42}
\end{equation}
with a Hamiltonian
\begin{equation}
H(s)=\frac{J}{4} \sum_{\bra x,x'\ket}s(x)s(x'),
\label{Hamil0}
\end{equation}
where $J=\log (2)$ and $\bra x,x'\ket $ represents 
a nearest neighbor pair $x$ and $x'$. 
Therefore,
the partition function  given in (\ref{eta-Xi}) is expressed as 
\begin{equation}
\Xi(\mu)=\e^{\frac{\mu}{2}N+\frac{J}{2}N}
\sum_{s \in \{-1,1\}^N}\e^{-H(s)-\frac{\mu}{2}\sum_x s(x)}.
\label{Xi-3}
\end{equation}
That is, $\Xi(\mu)$ is determined from the canonical partition 
function for the anti-ferromagnetic Ising model under a magnetic field.
Since $J/4=0.173..$ is less than the critical
point $\beta_{\rm c}\simeq 0.44...$ of the Ising model, 
the pressure function $p(\mu)$ does not exhibit 
any singularities as a function of $\mu$. Thus, there is no 
thermodynamic transition in this model. 
Related with the problem, we remark
that the expression (\ref{Xi-3}) with (\ref{Hamil0})
suggests the existence of 
a phase transition if $J$ were larger than $4 \beta_{\rm c}$.
Such a case  may arise in a $p$-prototiles model with 
sufficiently large $p$.
The transition in this case is regarded as an entropy-driven 
crystallization of holes. Although it is an interesting 
phenomenon, we do not discuss it in this paper.

% physics

Let us recall that there are an infinite series of 
complete tilings. When $\mu \to \infty$ is considered 
for fixed $L$, the expectation value of some observables
depends on the boundary conditions in the infinite size
limit. Note that this limit is different from the case 
$\mu \to \infty$ in the infinite system in which 
$L \to \infty$ is considered for the system with 
$\mu < \infty$. Since there is no 
thermodynamic transition, no boundary condition 
dependence of quantities  is observed in the latter case. 
That is, once holes are generated in a complete tiling 
with any positive ratio, the system becomes 
free from  boundary conditions despite the 
infinite degeneracy of complete tilings.
We may say that complete tilings in the infinite 
system are unstable with respect to  generation
of holes. The origin of the instability is understood 
from the fact that an infinite  region is influenced 
as the result of iteration of the cellular-automaton 
rule from one hole.  (Recall the argument above (\ref{basic}).)
Although a cellular-automaton rule can easily generate 
an infinite series of irregular complete tilings, 
it simultaneously leads to the instability of complete 
tilings so that a thermodynamic transition is not
observed.  In order to have stable complete tilings 
against generation of holes, we need to avoid 
a chain of tile-replacements induced by one hole.

%%%%%%%%%%%%%%%%%%%%%%%%%%%%%%%%%%%%%%%%%%%%%%%%%%%%%
\section{Stratified Wang tiles}\label{3dmodel}    %
%%%%%%%%%%%%%%%%%%%%%%%%%%%%%%%%%%%%%%%%%%%%%%%%%%%%%

% setting up

Let $\Lambda_{L,M} \equiv \{ (i,j,k) \in {\bf N}
\times {\bf N}\times {\bf N} | 1 \le i,j \le L, 1 \le k \le M \}$
be a cubic lattice.
As a natural extension of the  models in the previous section, 
we consider {\it  stratified Wang tiles} in $\Lambda_{L,M}$.
In addition to the color matching condition in each $(i,j)$ 
plane with $k$ fixed, we further impose  a constraint 
condition that two neighboring
prototiles in the $k$ direction, if they exists, are 
the same type. We define $\sigma(x)=n$ 
if there is a molecule congruent to a prototile $n \in C$ 
on the site $x \in \Lambda_{L,M}$ and as 
$\sigma(x)=0$ if the site $x$ is empty. Specifically,
we consider the 4-prototype model studied in the previous 
section.  A complete tiling in this model is given by 
$\sigma(i,j,k)=\sigma_\alpha(i,j)$ for $\alpha \in {\cal D}_\infty$.
We denote  it by $\sigma_\alpha(x)$ for $x \in \Lambda_{L,M}$.

% statistical mechanics

We study  statistical mechanics of the  model with a boundary 
condition imposed at sites in $\partial \Lambda_{L,M}^-$,
$\partial \Lambda_{L,M}^+$, and $\partial \Lambda_{L,M}^{\rm vert}$.
where 
$\partial \Lambda_{L,M}^-\equiv \{(i,j,k) \in \Lambda_{L,M} |i=1 \ {\rm
or}  \ j=1\} $,
$\partial \Lambda_{L,M}^+ \equiv \{(i,j,k) \in \Lambda_{L,M} |i=L \ {\rm
or } \ j=L\} $,
and $\partial \Lambda_{L,M}^{\rm vert} 
\equiv \{(i,j,k) \in \Lambda_{L,M} |k=1 \ { \rm or } \  k=M\} $.
The number of sites in the bulk region is $N=(L-2)^2(M-2)$. 
Without confusion, the notations in the two-dimensional case 
are also employed in the stratified model.
We assume that $\sigma$ obeys the grand canonical ensemble
\begin{equation}
P_\mu(\sigma)=\frac{1}{\Xi(\mu)}
D(\sigma)e^{\mu N \rho(\sigma)},
\label{gc-3d}
\end{equation}
where $D(\sigma)=1$ for configurations that satisfy the constraints
that contiguous edges of tiles in the $(i,j)$ plane for each $k$ 
have the same color and that neighboring prototypes in the $k$ direction
are the same type, otherwise $D(\sigma)=0$; 
$\mu$ is the chemical potential of molecules, 
and $\rho(\sigma)$ is the density of molecules defined by 
\begin{equation}
\rho(\sigma)=\frac{1}{N} \sum_{x \in \bar \Lambda_{L,M}}
[1-\delta(\sigma(x),0)].
\label{density-3d}
\end{equation}
The normalization constant $\Xi(\mu)$ in (\ref{gc-3d})
is the partition function of the stratified model.

\subsection{Glass transition}\label{glass}

%% thermodynamic transition/ definition of q

We  explain the existence of a thermodynamic transition
in this model. We fix a complete tiling 
$\alpha \in {\cal D}_\infty$.
We study statistical mechanics  (\ref{gc-3d}) of the
system  by assuming a boundary condition that
\begin{equation}
\sigma(x)=\sigma_\alpha(x)
\end{equation}
for $x \in \partial\Lambda_L^- $
and $\sigma(x)=0 $ for the other boundary sites.
We refer it to as  a {\it GS-boundary condition},
because the ground-state configuration is uniquely determined
by this condition.  The expectation value under the GS-boundary 
condition is denoted by $\bra \ \ket_\alpha^{L,M}$.
Here, let us recall that the up-spin boundary 
condition of the Ising model under which the ground state is
uniquely determined despite the system possesses the up-down 
symmetry. The order parameter of the Ising model is 
the magnetization and it is interpreted as the overlap with the
ground-state configuration. Similarly, in our model,
as an order parameter associated with the complete tiling $\alpha$, 
we define  an overlap variable with $\alpha$ as 
\begin{equation}
q_\alpha(\sigma) \equiv \frac{1}{N}
\sum_{x \in \bar \Lambda_{L,M}}
\delta(\sigma(x), \sigma_\alpha(x)).
\end{equation}

%% q under the GS-boundary

Now, we consider the case that  $1  \ll \e^{\mu} \ll L=M$. 
A typical configuration in this case is given by a random 
deposition of holes with a probability with $e^{-\mu}+O(\e^{-2\mu})$
for each site. When a hole is inserted to the complete tiling,
no other tile configurations are allowed,
which is different from the two-dimensional case discussed
in the previous section. A tile different from $\sigma_\alpha$
at a site $x_0$ appears when four holes enclose
the site $x_0$. The probability of such a configuration
is $2 \e^{-4\mu}+O(\e^{-5\mu})$. In general, we expect that 
$\lim_{L \to \infty} \bra q_\alpha  \ket_\alpha^{L,L}$
is expressed as a convergent expansion in $\e^{-\mu}$
for sufficiently large $\mu$.  That is, 
\begin{equation}
\lim_{L \to \infty} \bra q_\alpha  \ket_\alpha^{L,L}
= 1-e^{-\mu}+O(e^{-2\mu}).
\label{op-GS}
\end{equation}
for large $\mu$.  A more precise estimation on the basis
of a Peierls argument should be presented.
(See a brief discussion in section \ref{remarks}.)

%% q under the open-boundary

Next, we assume  the open boundary condition
that $\sigma(x)=0 $ for all the  boundary sites.
We denote the expectation value under the boundary condition 
by $\bra \ \ket_{0}^{L,L}$.  Then, when $\mu$ is sufficiently
large, a typical configuration is close a complete tiling,
but it is not identical to $q_{\alpha}$, in general. 
As discussed in section \ref{complete}, 
the overlap ratio between two different complete 
tilings  is $1/4$. We  thus obtain
\begin{equation}
\lim_{L \to \infty}\bra q_\alpha \ket_{0}^{L,L} \simeq \frac{1}{4}
\label{op-open}
\end{equation}
for sufficiently large $\mu$. By comparing (\ref{op-GS}) and 
(\ref{op-open}), we conclude that the expectation value of the 
observable $q_{\alpha}$ depends on boundary conditions 
in the infinite size limit. 
This means that there exists the ordered phase in the system with 
sufficiently large $\mu$. In the other limit where $\mu$ is 
sufficiently small (negatively large), 
statistical properties are independent of boundary conditions,
because dilute tiles are almost non-interacting.
Thus, there exists a 
thermodynamic transition at some value of $\mu$. 

%% order parameter 

The quantity characterizing the ordered phase is
a collection  of $q_{\alpha}$,
which is denoted by $\bv{q}=(q_{\alpha})_{\alpha \in {\cal D}_\infty}$.
Formally, the order parameter $\bv{q}$ is  an uncountably 
infinite dimensional vector. 
Such an order parameter is quite peculiar. To our best knowledge, 
no examples in finite dimensions have  been reported.
We also recall that a typical complete tiling $\alpha$ 
does not exhibit Bragg peaks in its Fourier transform.
From these conditions, we 
identify the thermodynamic transition to the ordered phase
with  the glass  transition to the glass phase characterized
by $\bv{q}$. 

\subsection{Numerical experiments}\label{n-exp}

% caution on the numerical experiments

We  report results of numerical experiments 
for the stratified 4-prototiles model. 
In order to facilitate the equilibration, we employ
the exchange Monte Carlo method \cite{HukushimaNemoto}.
We prepare $K$ replicas of the system with $\mu_k=\mu_{\rm max}k/K$,
$k=1,\cdots,K$. In this paper, we set $\mu_{\rm max}=2$ and $K=40$.
We estimate the expectation value $\bra A \ket_{\alpha}^{L,M}$ 
for an observable $A(\sigma)$ by the time average of $A(\sigma(t))$
during the time interval $[t_{\rm w}, 2 t_{\rm w}]$ with 
discarding the transient data $[0,t_{\rm w}]$, 
where the initial condition was assumed to be 
$\sigma=\sigma_\alpha$ for all $\mu_k$. 
When the time average of $A(\sigma(t))$ is independent of
$t_{\rm w}$ within statistical errors, we assume that the 
result provides the estimation of  equilibrium values 
$\bra A \ket_{\alpha}^{L,M}$.

%%%% figure (density ) %%%%%

\begin{figure}[htbp]
\begin{center}
\includegraphics[width=6cm]{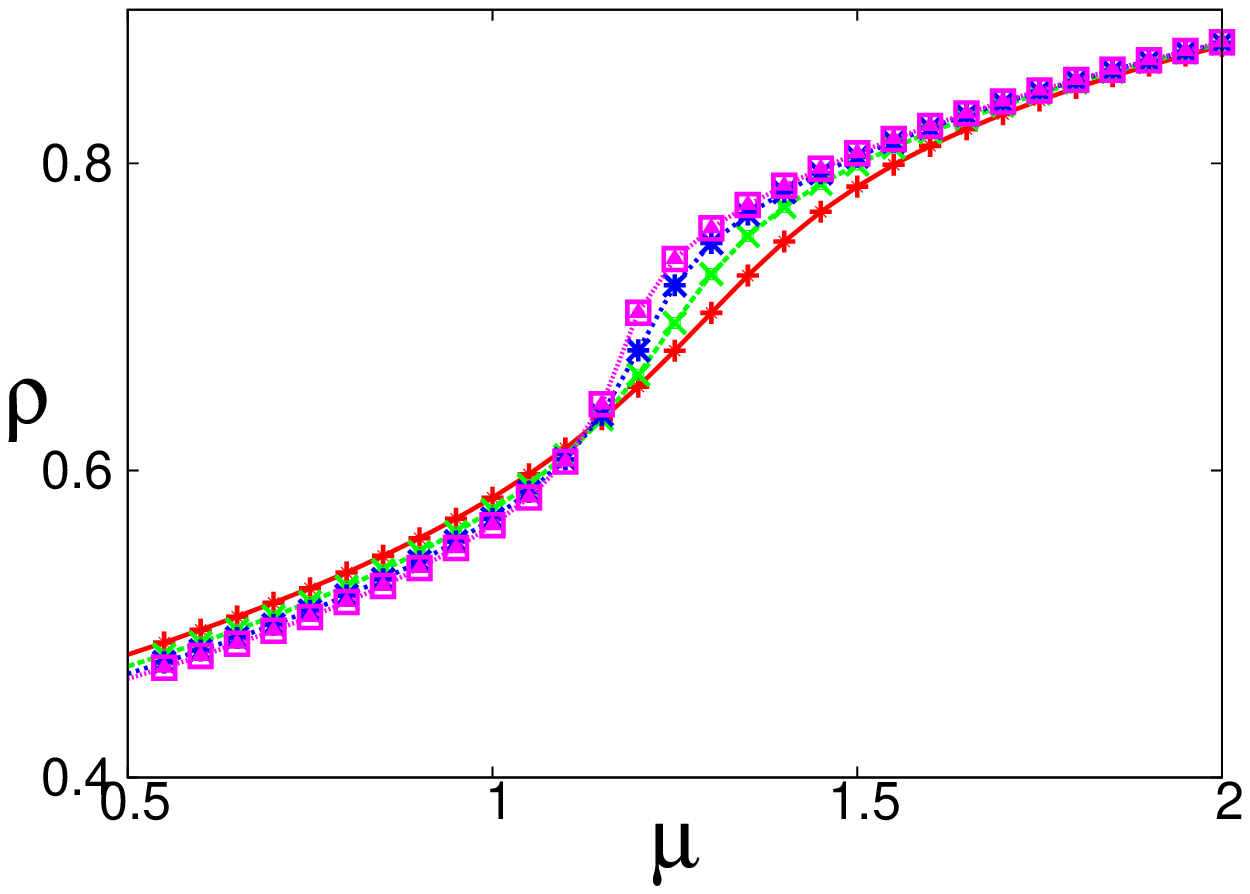}
\includegraphics[width=6cm]{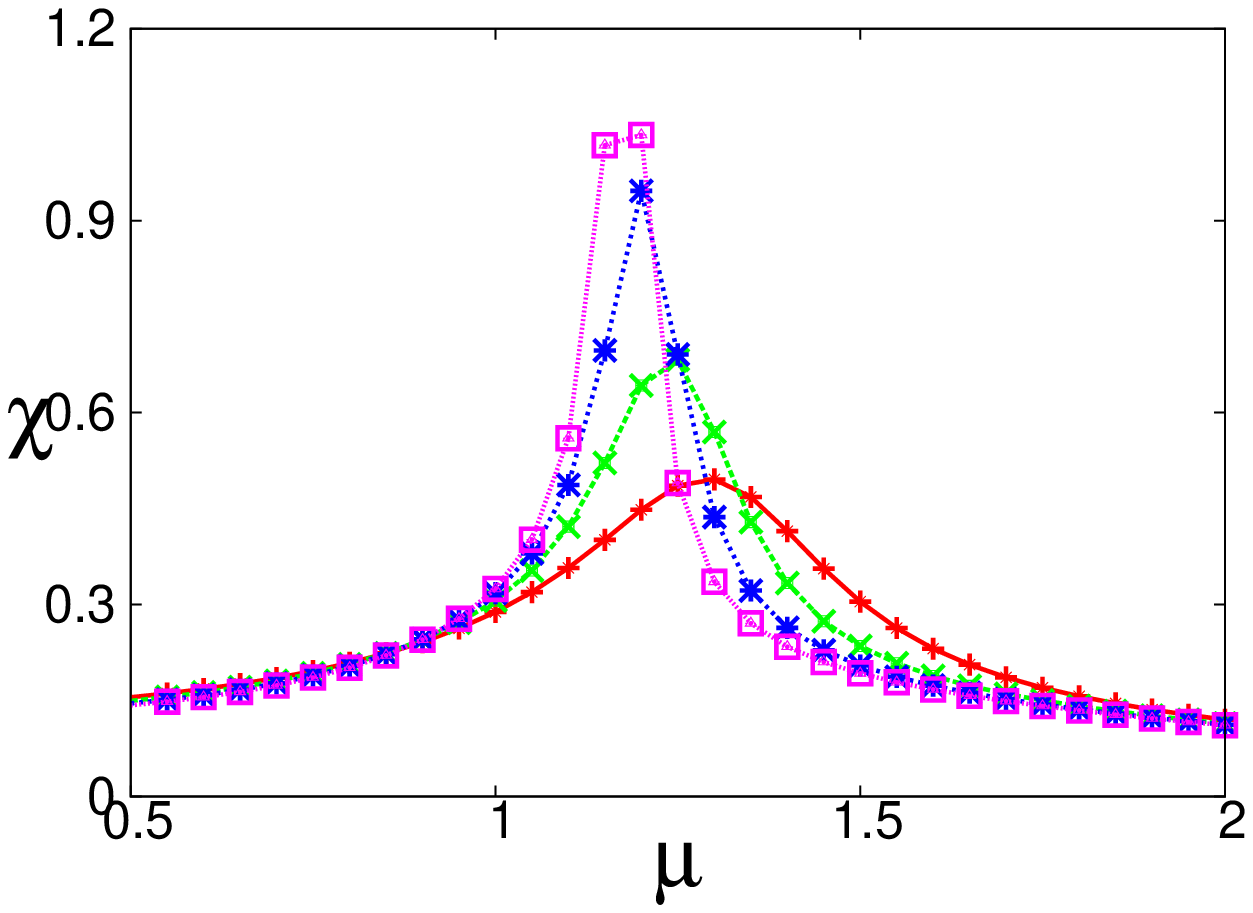}
\end{center}
\caption{Statistical quantities of the density.
$(L,M)=(12,8)$ (plus), $(15,10)$ (cross), $(18,12)$ (asterisk), 
and $(21,14)$ (box).
Error-bars are within the symbols.
Left: Expectation value of the density. 
Right: Fluctuation intensity of density $\chi$.
}
\label{den}
\end{figure}
%%%%%%%%%%%%%%%%%%%%%%%%%%%%%%%

% density 

In the left side of figure \ref{den}, 
$\bar \rho\equiv \bra \rho \ket_{\alpha}^{L,M}$
is plotted as a function of $\mu$ for the system  with four different
system sizes. In order to clarify the singular nature, 
the fluctuation intensity 
defined by $\chi \equiv N \bra (\rho- \bar \rho)^2 \ket_{\alpha}^{L,M}$
is also displayed in the right side of figure \ref{den}.
It should be noted that the fluctuation relation 
$\chi= d\bar \rho(\mu)/d\mu$ holds.  The graphs of $\chi$
indicate the existence of a thermodynamic transition. 

%%%% figure (glass order parameter q) %%%%%

\begin{figure}[htbp]
\begin{center}
\includegraphics[width=6cm]{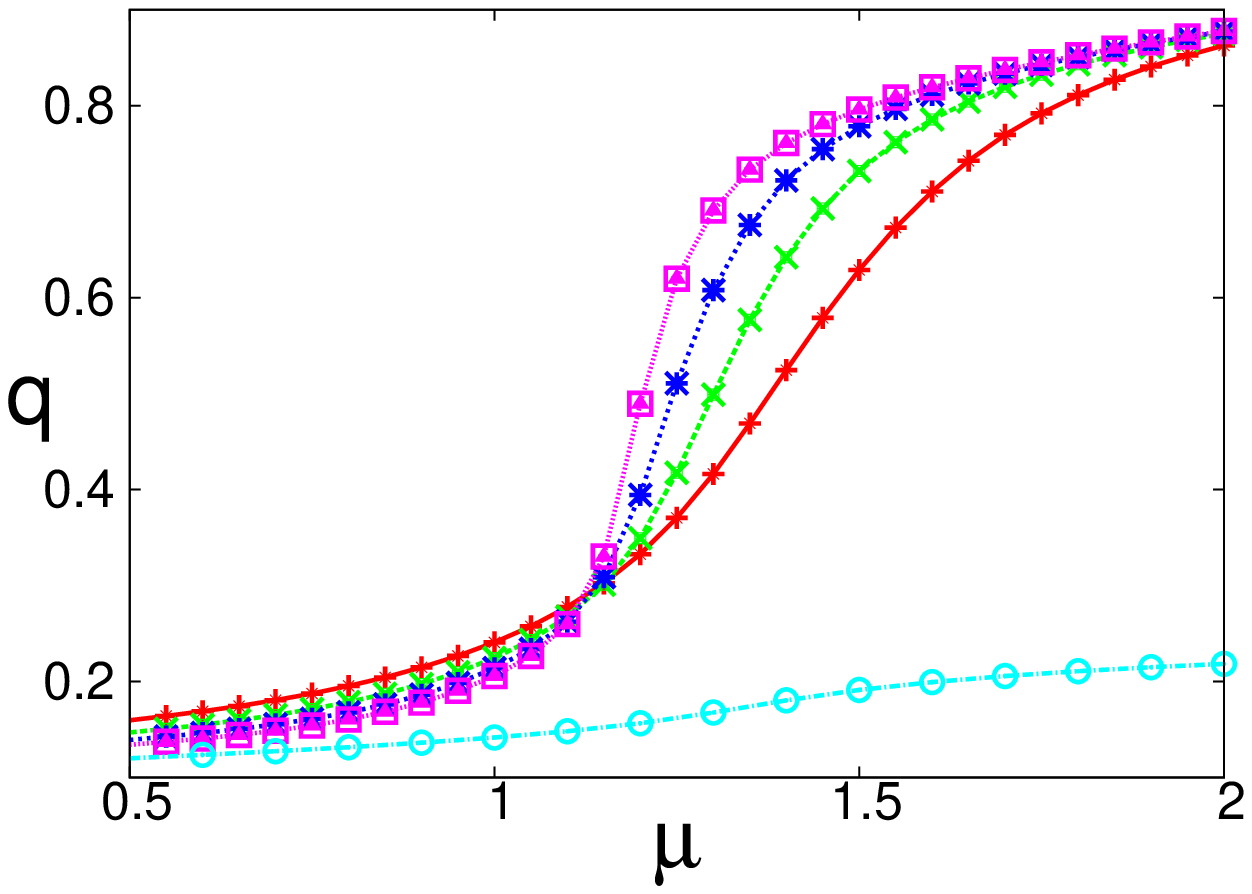}
\includegraphics[width=6cm]{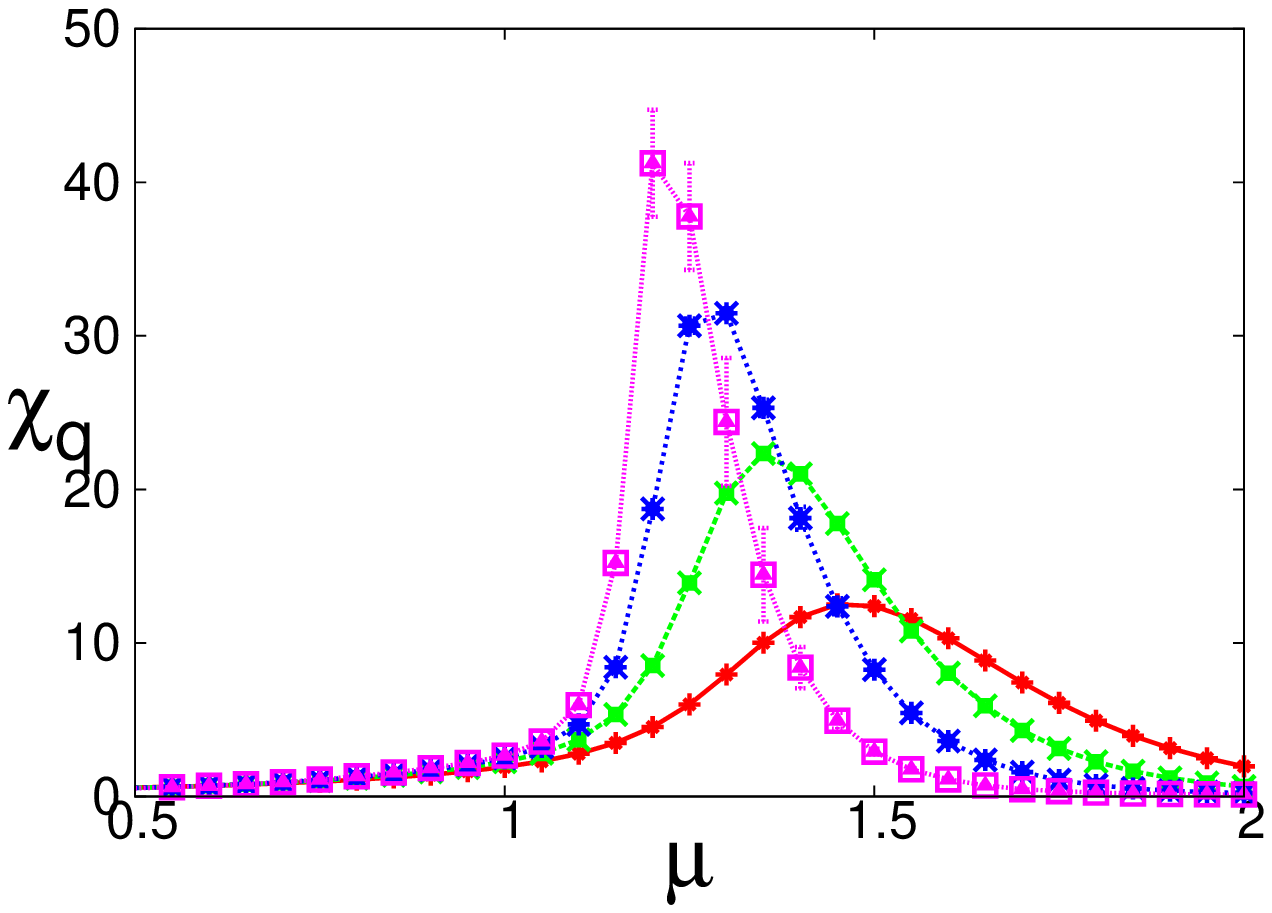}
\end{center}
\caption{Statistical quantities of the glass order parameter. 
$(L,M)=(12,8)$ (plus), $(15,10)$ (cross), $(18,12)$ (asterisk), 
and $(21,14)$ (box). Error-bars are within the symbols
expect for $\chi_q$ with $(L,M)=(21,14)$.
Left: Expectation value 
$\bra q_{\alpha} \ket_{\alpha}^{L,M}$
and $\bra q_{\alpha} \ket_{0}^{12,8}$ (circle).
Right: Fluctuation intensity of the order parameter
$\chi_q$.
}
\label{q-order}
\end{figure}
%%%%%%%%%%%%%%%%%%%%%%%%%%%%%%%

%% order-parameter 

Next, we  numerically investigate the existence of 
the glass phase by measuring the order parameter $q_{\alpha}$. 
In the left side of figure \ref{q-order}, 
we display $\bar q \equiv \bra q_{\alpha} \ket_{\alpha}^{L,M}$
and $\bra q_{\alpha} \ket_{0}^{L,M}$, which shows
the boundary condition dependence of the expectation
value of $q$. From the size dependence, we expect 
that the behavior 
sustains in the infinite size limit.
This means that the transition is identified with the glass 
transition. Note that the expectation value of 
the density is independent of boundary conditions in the thermodynamic
limit.
In order to characterize the singularity associated with the 
order parameter, we  measured 
$\chi_q\equiv N \bra (q_\alpha- \bar q )^2 \ket_{\alpha}^{L,M}$.
As shown in the right side  of figure \ref{q-order}, $\chi_q$ 
exhibits a divergent behavior at some value of $\mu$.

%%%% figure (power-law) %%%%%

\begin{figure}[htbp]
\begin{center}
\includegraphics[width=6cm]{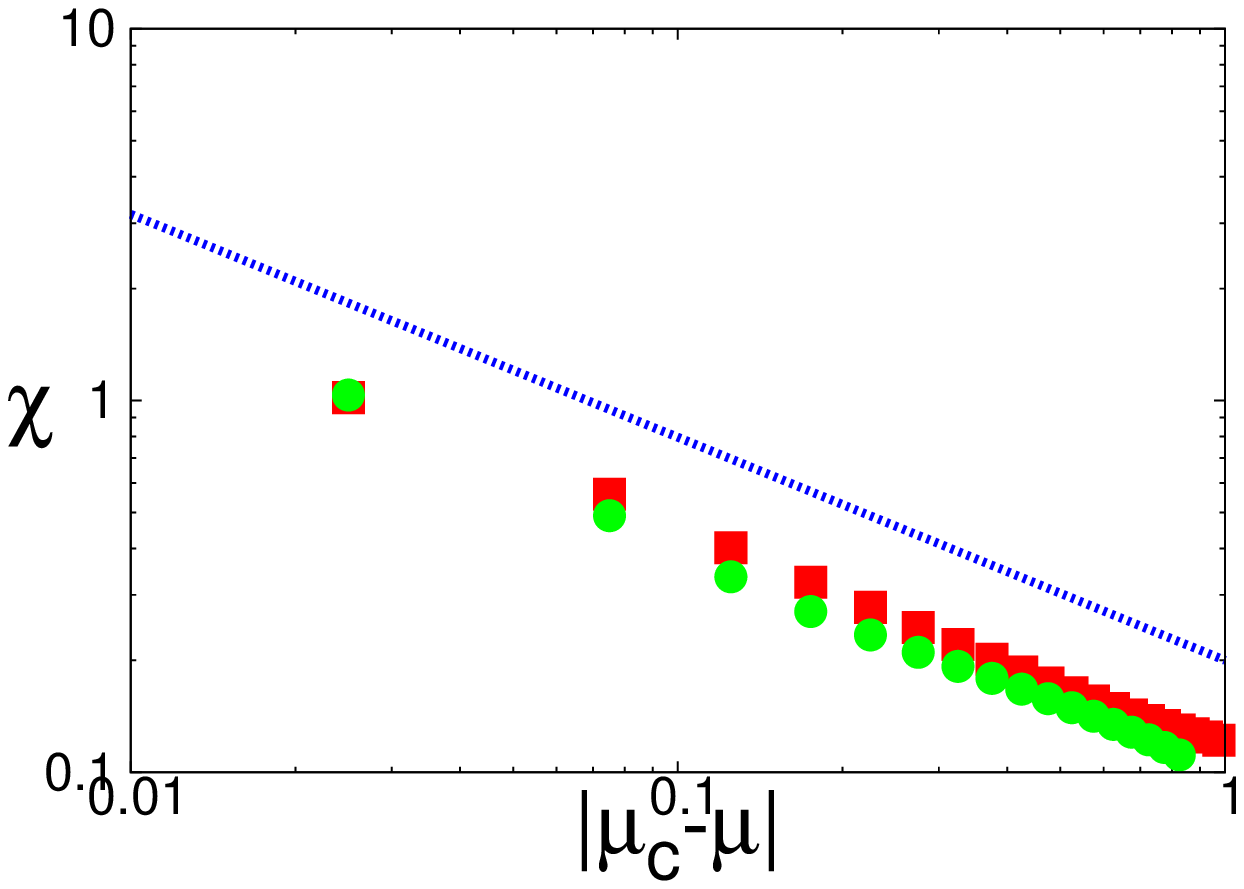}
\includegraphics[width=6cm]{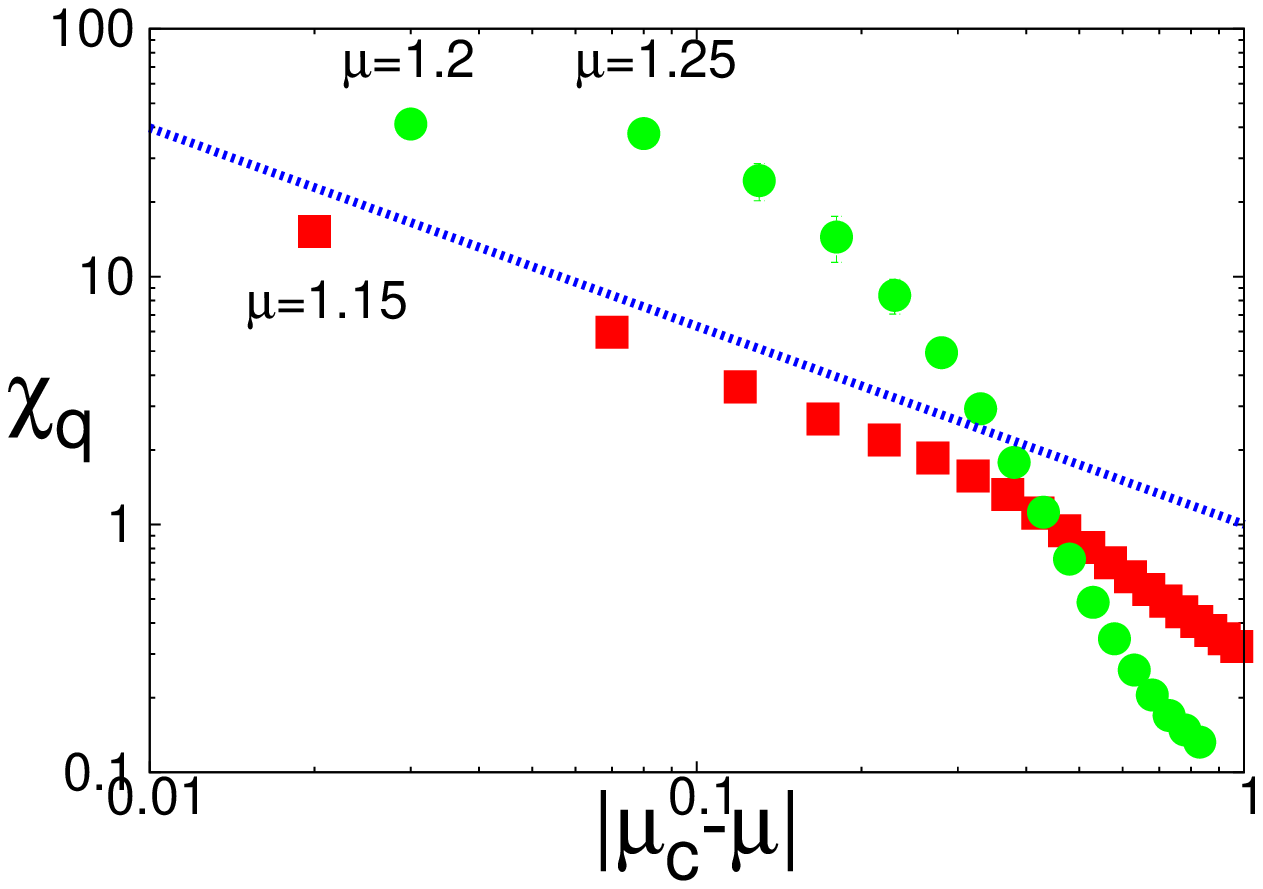}
\end{center}
\caption{
Left: Log-log  plot of $\chi$ as a function of 
$|\muc-\mu|$, where $\muc$ is assumed to be $1.175$,
for $(L,M)=(21,14)$. The square symbols correspond to 
the case $\mu < \muc$
and the circle symbols to $\mu > \muc$. 
The guide line represents 
$\chi=0.2|(\muc-\mu)|^{-0.6}$.
Right: Log-log  plot of $\chi_q$ as a function of 
$|\muc-\mu|$, where $\muc$ is assumed to be $1.17$,
for $(L,M)=(21,14)$.
The square symbols correspond to the case $\mu < \muc$
and the circle symbols to $\mu > \muc $. The guide line represents 
$\chi_q=(\muc-\mu)^{-0.8}$.
}
\label{power-law}
\end{figure}

%%%%%%%%%%%%%%%%%%%%%%%%%%%%%%%%%%

%% power-law divergence ?

Here, we study the nature of the singularity quantitatively.  
We first attempt to fit the fluctuation intensities
$\chi$ and $\chi_q$  with power-law functions as $|\mu-\muc|$,
where $\muc$ is a transition point.
As displayed in the left side of figure \ref{power-law}, 
$\chi$ may exhibit a power-law behavior 
\begin{equation}
\chi \simeq |\mu-\muc|^{-\alpha},
\end{equation}
with $\alpha \simeq 0.6$ and $\muc = 1.175$.
This result indicates that the transition is the second order 
according to the Ehrenfest classification.
On the other hand, 
a clear power-law behavior is not observed in $\chi_q$. 
As shown in the right side of figure \ref{power-law},
a fitting of the power-law form $\chi_q \simeq (\muc-\mu)^{-\gamma}$ 
might be not so bad with $\gamma \simeq 0.8$
when  we choose  a  value of $\muc$ ($\muc=1.17$). 
However,  $\chi_q$ in the regime $\mu >\muc$ is far from 
power-law behaviors even if we change the value of $\muc$.
The singular nature of order parameter fluctuations 
is quite unusual.

%%%% figure (distribution) %%%%%

\begin{figure}[htbp]
\begin{center}
\includegraphics[width=7cm]{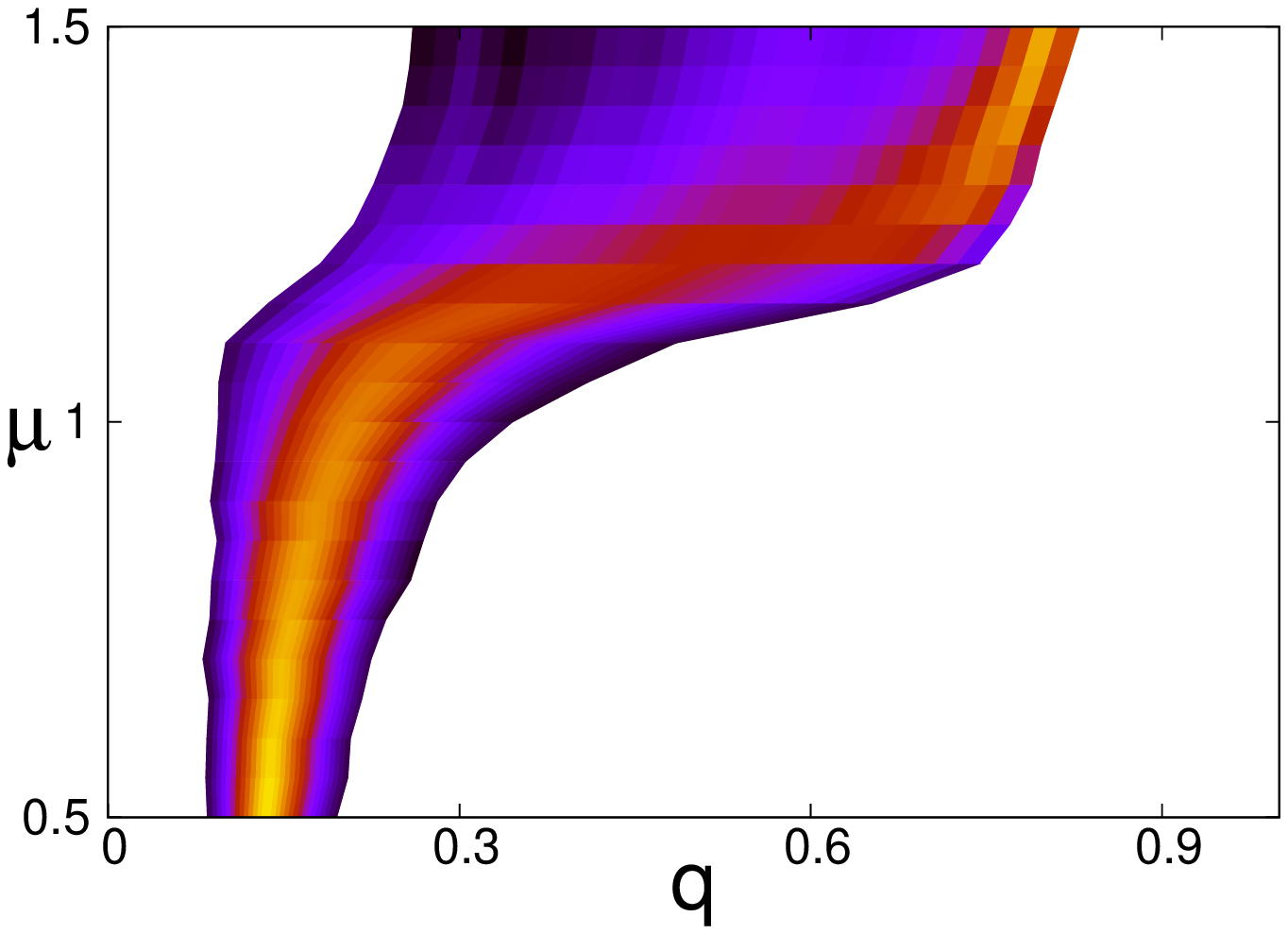}
\includegraphics[width=7cm]{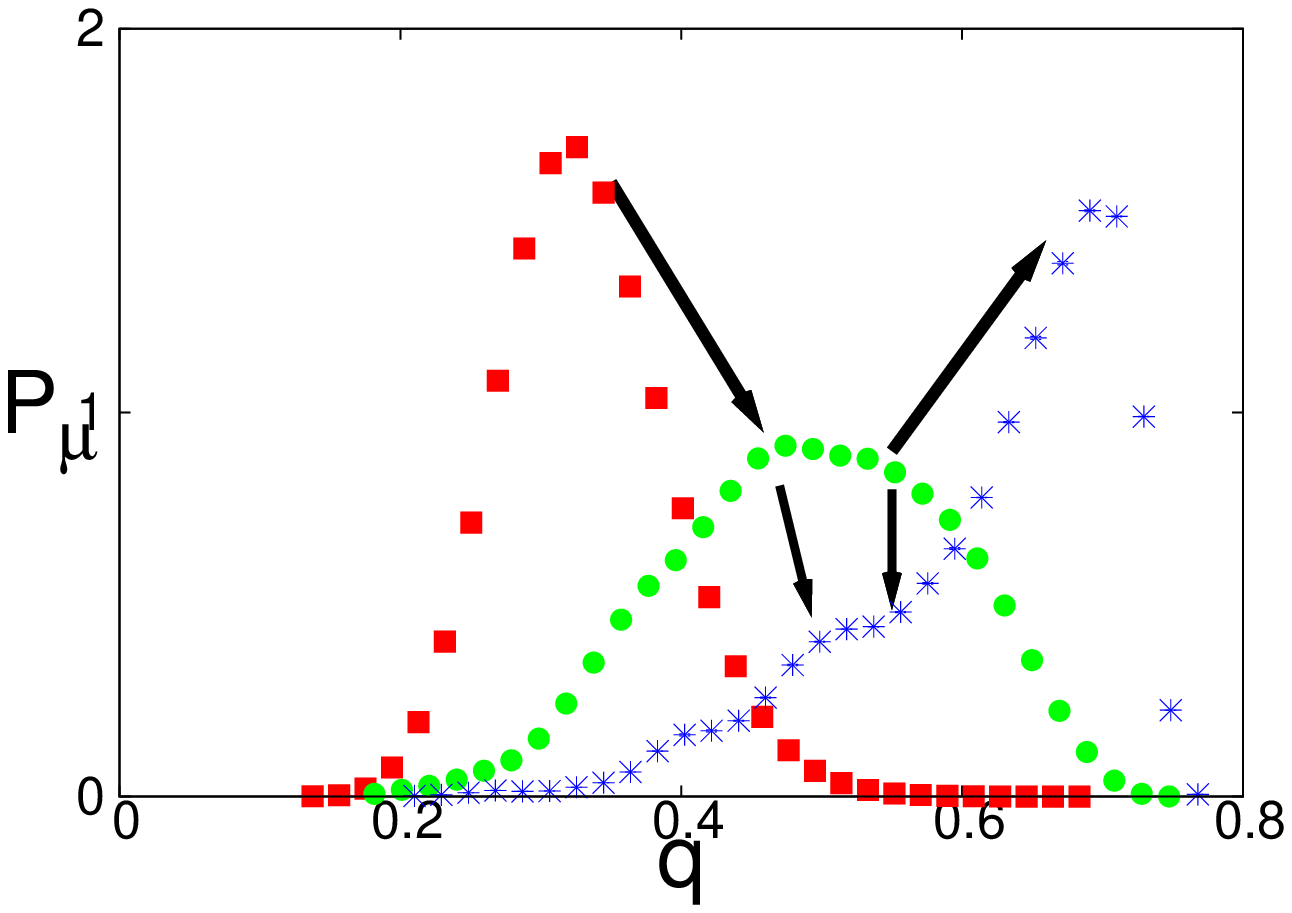}
\end{center}
\caption{Left: Color representation of
$P_\mu(q)$ in the $(q,\mu)$ space. $(L,M)=(21,14)$.
Right: $P_\mu(q)$ with $\mu=1.15$ (square),
$\mu=1.2$ (circle), and $1.25$ (asterisk). 
The error-bars are within the symbols. 
}
\label{dist}
\end{figure}
%%%%%%%%%%%%%%%%%%%%%%%%%%%%%%%

%%%

In order to demonstrate the peculiar behavior more, 
we consider the distribution function of $q$
for the system with $\mu$, which is denoted by
$P_\mu(q)$. The left side of figure \ref{dist} 
shows that  the peak of $P_\mu(q)$ for $\mu > \muc$
accompanies a broad tail in the smaller $q$ region.
We investigate $P_\mu(q)$ near a transition point
carefully in the right side of figure \ref{dist}.
Let us focus on the graph with $\mu=1.25$ represented
by the asterisk symbols, which corresponds to the 
second left of circle symbols in the right of figure
\ref{power-law}. 
There is a small slope regime 
around $q=0.5$ which is apart from the main peak 
$q=0.7$. The left edges of the regime  seems to
be connected to the left edge of the nearly flat regime 
of the graph with $\mu=1.2$, which corresponds to the
most left of circle symbols in  the right of figure
\ref{power-law}. 
The left edge of the flat
regime may be further connected to the main peak of
the graph with $\mu=1.15$, 
which corresponds to the
most left of square  symbols in  the right of figure
\ref{power-law}. 
On the other hand,
the main peak of the graph with $\mu=1.25$
arises from the right edge of the flat regime of 
the graph with $\mu=1.2$. The important
thing here is that the main peak in the ordered
phase is not connected to the main peak in the
disordered phase. 
This suggests
the discontinuous transition of $\bra q \ket_\alpha^{L,M}$
in the thermodynamic limit $L,M \to \infty$,
although a clear 
discontinuous jump is not observed in the left side 
of figure \ref{q-order}. 
Extensive numerical studies are 
necessary to have firm  evidences for supporting
the conjecture, because the sizes we investigated
are too small. In any cases, we can say that the behavior
of the order parameter $\bv{q}$ is quite unusual.

%%%%%%%%%%%%%%%%%%%%%%%%%%%%%%%%%%%%%%%%%%%%%%%%
\section{Concluding remarks} \label{remarks}  %
%%%%%%%%%%%%%%%%%%%%%%%%%%%%%%%%%%%%%%%%%%%%%%%%

% opening

In this paper, we have proposed lattice molecule models, 
one of which exhibits the thermodynamic glass transition
in three dimensions. The glass phase is characterized by 
the uncountably-infinite dimensional order 
parameter $\bv{q}$, each component of which measures 
the overlap with an irregular ground state. 
We conjecture that the transition is the second-order 
in the sense of thermodynamics, while the glass order 
parameter exhibits a unusual behavior. 
Before ending the paper, we describe problems that 
should be studied seriously in future.  

% Other Wang tiles

Although  we have studied the system with the simplest 
set of 4-prototiles that generates irregular complete tilings,
statistical behavior  of lattice molecule models
depends on the choice of  prototiles. By studying  models
with  other  sets of prototiles, we wish to  classify their 
phenomena. 
In particular, it is interesting to find a two-dimensional model 
that exhibits a glass transition or to prove that there is no 
such a model. Note that the thermodynamic transition to the 
quasi-periodic ordered phase was observed in a two-dimensional 
Wang-tiles model \cite{Leuzzi,Radin}. The difference between 
the quasi-periodic order and the glass order should be clarified. 
In doing these studies, extensive numerical experiments are 
necessary. It is significant to develop an efficient method
for numerical calculation. The hard nature of molecules 
would  substantially reduce the computation
time if an elegant algorithm is found.

% theory-1

\begin{figure}[htbp]
\begin{center}
\includegraphics[width=6cm]{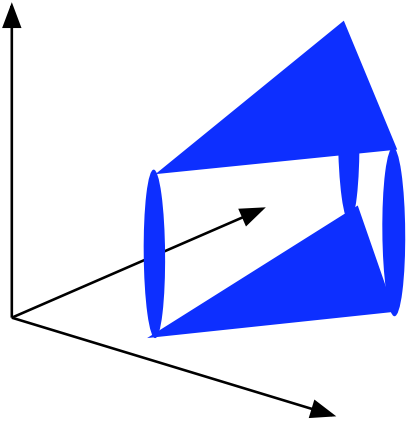}
\end{center}
\caption{Schematic figure of a simple 
interface configuration with large $\eta$.
Holes in the top and bottom triangles form Sierpinski 
gaskets and the vertexes of the triangles are connected by 
holes in the vertical directions. 
}
\label{wall}
\end{figure} 

Since the theoretical arguments reported in this paper are 
still in  the early stage of the study, important theoretical 
problems remain to be solved. The first problem  is to provide 
a mathematical proof 
of the existence of the glass transition. This might 
be solved as follows. We consider a probability that 
$\sigma$ at the center site $x_0$ is different from
$\sigma_\alpha$ under the boundary condition $\sigma=\sigma_\alpha$
for all boundary sites. We estimate an upper-bound of 
the probability by noting
an {\it interface} separating an ordered region connected to the boundary
sites. (See an example of an interface in figure \ref{wall}.)
If this is explicitly  defined for a given arbitrary 
configuration 
in which $\sigma(x_0) \not = \sigma_\alpha(x_0) $, we may formulate 
a Peierls argument. That is, a transition is understood
from the competition between the entropy cost of configurations
with interfaces  and the energy cost of holes. However, up to
the present, we do not have an explicit definition of such interfaces.

% theory-2

A more important but difficult problem is to obtain a 
mathematical description of statistical properties
of the glass order parameter $\bv{q}$. When we consider this 
problem, it seems better to forget the tile model. Instead, 
we will analyze  a stratified model of the 3-body spin 
interaction on  upward triangles.  (See a remark at the end
of section \ref{complete}.)
By applying the same argument 
in the present paper to the spin model, we may find a 
glass transition for the stratified model. Since 
this model is much simpler than the tile model,
several theoretical calculations including a
sort of the Bethe approximation will be done
more easily. Such theoretical study  may provide
a connection to the mean-field picture of the 
glass transition (RFOT). 
Furthermore, it is  amazing if we find  an exactly solved 
model that exhibits a glass transition. An exact solved 
model would play a crucial influence on the study of glass transitions.

% math-phys

In the analysis of infinite-range interaction models that exhibit
a glass transition, the number of pure states in the glass phase 
has been one of concerns.
With regard to this problem, we briefly review  
Gibbs measures of infinite-size systems. 
Roughly speaking, a Gibbs measure is defined in such a way that the 
probability of configurations in any finite-size region 
is given  by the grand canonical ensemble with boundary 
conditions which are  chosen by the measure. In the disordered 
phase, the Gibbs measure is unique, while there are an infinite 
number of Gibbs measures if the uniqueness is broken. 
In particular, a special measure that cannot be decomposed 
further into a superposition of other Gibbs measures
is called  a pure state. Here, let us recall that
each GS-boundary condition in the Ising model can provide
a pure state. Although GS-boundary conditions 
are not always related to pure states, the statistical
ensembles in finite systems under GS-boundary conditions
may be a starting point for understanding
of pure states in the glass phase. 

% glass physics

Finally, we go back to our motivation of understanding 
the nature of glassy materials. Although we have 
found a thermodynamic glass transition in a short-range
interaction model in finite dimensions,
it is not obvious whether or not such an idealized 
phase is actually observed in laboratory experiments. 
Toward an experimental realization of thermodynamic 
glass phases, we need to consider the following problems. 
The first is to construct a mechanical model that exhibits 
a glass transition. Although we have only to design a 
potential function sharing common features with hard-constraint 
conditions in Wang tiles, its explicit demonstration may be  
challenging. The second problem is to find  an experimentally 
realizable algorithm for facilitating the equilibration,
because the exchange MC method cannot be employed in 
laboratory experiments. After solving the problems,
we hope that we will be able to demonstrate by numerical 
experiments that a genuine 
thermodynamic glass transition is observed in 
laboratories.

%%%%%%%%%%%%%%%%%%%%%%%%%%%%%%%%%%%%%
% acknowledgment                    %
%%%%%%%%%%%%%%%%%%%%%%%%%%%%%%%%%%%%%

\ack

The author thanks K. Hukushima, H. Ohta, and H. Tasaki 
for continuous 
discussions on statistical mechanics of glassy systems. 
He  also thanks T. Chawanya for communications toward
understanding of Ref. \cite{Kari}.  
The present study was supported by grants from the 
Ministry of Education, Culture, Sports, Science, and Technology 
of Japan, Nos. 22340109 and 23654130. 

\appendix

\section*{References}

%\section*{References}


\begin{thebibliography}{99}% 

%%% glasses

% RFOT
\bibitem{Cavagna}
Cavagna A 2009  Physics Reports {\bf 476}, 51

\bibitem{GibbsDiMarzio}
Gibbs J H and DiMarzio E A 1958 
J. Chem. Phys. {\bf 28}, 373 

\bibitem{AdamGibbs}
Adam G and  Gibbs J H  1965 
J. Chem. Phys. {\bf 43}, 139

\bibitem{KTW}
Kirkpatrick T R, Thirumalai D, and  Wolynes P G,
1989 Phys. Rev. A {\bf 40},  1045

\bibitem{BB:JCP}
Bouchaud J P and  Biroli G 2004
J. Chem. Phys. {\bf 121} 7347

% statistical mechanics of glass transition

\bibitem{Monasson}
Monasson R  1995  Phys. Rev. Lett., \textbf{75}, 2847

\bibitem{MezardParisi}
M\'ezard M and  Parisi G  1999  Phys. Rev. Lett. {\bf 82} 747

\bibitem{BiroliMezard}
Biroli G and  M\'ezard M 2002  Phys. Rev. Lett. {\bf 88}, 025501

\bibitem{Rivoire}
Rivoire O et al 2004 Eur. Phys. J. B {\bf 37}, 55 

\bibitem{Krzakala} 
Krzakala F, Tarzia M, Zdeborov\'a L 2007 
Phys. Rev. Lett. {\bf 101}, 165702

\bibitem{ParisiZamponi}
Parisi G and Zamponi F  2010 Rev. Mod. Phys. {\bf 82} 789

%\bibitem{MezardParisi}
%Mezard M and Parisi G 2009 arXiv:0910.2838

%%% van der Waals ??

\bibitem{Gallavotti}
Gallavotti 1999  {\it Statistical Mechanics  A Short Treatise}
(Springer, Berlin)

%% pionnear of phase transition in finite dimension

\bibitem{Peierls}
Peierls R 1936  Proceesdings of Cambridge Philosphical Society
{\bf 32} 477

\bibitem{KrameresWannier}
Kramers H A and Wannier G H 1941 Phys. Rev. {\bf 60} 252

\bibitem{Onsager}
Onsager L 1944 Phys. Rev. {\bf 65} 117 

%\bibitem{history-on-RG}
%A textbook on the histrory of RG 

%%% status of glass in finite dimensions

%\bibitem{BMcrystal}
%refereces on crystals on BM 

\bibitem{HukushimaSasa}
Hukushima K and Sasa S 2010  
J. Phys.: Conf. Ser. {\bf 233} 012004 

%%% parisi

\bibitem{LatticeGlass}
Pica Ciamarra M,  Tarzia M,  de Candia A, and  Coniglio A
2003 Phys. Rev. E. {\bf 67} 057105

\bibitem{Parisi}
Parisi G 2009 arXiv:0911.2265

\bibitem{disorder-spin}
Brangian C, Kob W, and Binder K 2002
J. Phys. A: Math.Gen. {\bf 35} 191


\bibitem{Moore}
Moore  M A 2006 
Phys. Rev. Lett.  {\bf 96} 137202


% Wang  tiles

\bibitem{Tiling}
Gr\"unbaum B and Shephard G C 1987 
{\it Tilings and Patterns} (W. H. Freeman and Company, New York)

\bibitem{Robinson}
Robinson R M 1971 Inventions Math. {\bf 12} 177 

\bibitem{Davis}
Davis M  1982 {\it  Computability and Unsolvability} (Dover, New York)

\bibitem{Culik}
Culik II, K 1996
Discrete Mathematics {\bf 160} 245

\bibitem{Leuzzi}
Leuzzi L and Parisi G 2000 J. Phys. A:Math. Gen. {\bf 33} 4215

\bibitem{Radin} 
Aristoff D and Radin C 2011 arXiv:1102.1982


\bibitem{Kari}
Kari J 1996
Discrete Mathematics {\bf 160} 259 

\bibitem{cha}
The author learned this fact from T. Chawanya. 

% dynamical systems
\bibitem{DS}
Powell G E and Percival I C 1979 J. Phys. A: Math. Gen, {\bf 12} 2053

\bibitem{ER}
Eckmann J P and Ruelle D 1985 Rev. Mod. Phys. {\bf 57} 617

% quasi-crystal

\bibitem{QC}
Janssen T  1988 Physics Report {\bf 168} 55


%Enter
\bibitem{Enter1}
van Enter A C D, Miekisz J
1992 J. Stat. Phys. {\bf 66} 1147


\bibitem{Enter2}
van Enter A C D, Miekisz K, and Zahradnik M 
1998 J. Stat. Phys. {\bf 90} 1441

\bibitem{CA}
Wolfram S  1983 Rev. Mod. Phys. {\bf 55} 601

% upward triangle Model

\bibitem{NM}
Newman M E J and Moore C 1999 Phys. Rev. E {\bf 60} 5068

\bibitem{GN}
Garrahan J P and Newman M E J 2000 Phys. Rev. E {\bf 62} 7670
 
\bibitem{JG}
Jack R and Garrahan J P 2005 J. Chem. Phys {\bf 123} 164508

\bibitem{Sasa}
Sasa S 2010 J. Phys. A:  Math. Theor. {\bf 43} 465002


% exchange MC

\bibitem{HukushimaNemoto}
Hukushima K and Nemoto K  1996
J. Phys. Soc. Jpn. \textbf{65} 1604

%\bibitem{pure}
%reference on pure states


\end{thebibliography}
\end{document}